\documentclass[prl,aps,twocolumn,showpacs]{revtex4}
\usepackage{graphicx}
\newcommand{\nn}{\nonumber}

\begin{document}

\title{Beliaev Damping and Kelvin Mode Spectroscopy of a Condensate \\ in the Presence of a Vortex Line}

\author{T. Mizushima}
%\email{mizushima@mp.okayama-u.ac.jp}
\affiliation{Department of Physics, Okayama University,
             Okayama 700-8530, Japan}
\author{M. Ichioka}
\affiliation{Department of Physics, Okayama University,
             Okayama 700-8530, Japan}			 
\author{K. Machida}
\affiliation{Department of Physics, Okayama University,
             Okayama 700-8530, Japan}
\date{\today}

%>>>########################### ABSTRACT ############################## 
\begin{abstract}
It is demonstrated theoretically that the counter-rotating quadrupole mode 
in a vortex of Bose-Einstein condensates
can decay into a pair of Kelvin modes via Beliaev process.
We calculate the spectral weight of density-response function within 
Bogoliubov framework, taking account of both Beliaev and Landau processes.
Good agreement with experiment on $^{87}$Rb by Bretin {\it et al}. [cond-mat/0211101]
allows us to unambigiously identify the decayed mode as the Kelvin wave propagating along a vortex line.
\end{abstract}
%<<<########################### ABSTRACT ##############################

\pacs{03.75.Fi, 05.30.Jp, 67.40.Vs}

\maketitle

Much attention has been focused on Bose-Einstein condensation (BEC)
realized in alkali-atom gases \cite{dalfovoRMP}.
Quantized vortices which are a hallmark of superfluidity \cite{fetterJP} are created in various methods,
such as phase imprinting \cite{matthews}, mechanical rotation by optical spoon \cite{madison} and 
topological Berry phase engineering \cite{leanhardt}.
Hundreds of vortices are trapped in a BEC system \cite{abo01}.
The creation and decay processes of vortices are investigated 
experimentally and theoretically, 
giving rise to a general consensus \cite{fetterJP} that quantized vortices in a scalar,
i.e. one-component BEC are well described by 
the Bogoliubov framework regarding to the static properties, such as 
the density profile or core radius, etc.

In contrast, regarding to the dynamical aspects, 
the study of low-lying collective modes is rather scarce 
in theory and particularly in experiment.
Needless to say, the low-lying Fermionic excitations in a vortex have played a 
fundamental role in charged or neutral Fermion systems, 
that is, the mixed state in a superconductor \cite{sc} and 
superfluid $^3$He.
Here we have a unique opportunity to investigate Bosonic excitations associated with 
a vortex, which was difficult in superconductivity.
In particular the so-called Kelvin mode \cite{donnelly} 
propagating along the vortex line, which is studied in classical normal fluids and 
superfluid $^4$He, 
is interesting to identify and characterize in the present dilute Bose gases.
This is an unexplored region.

Note that Bosonic excitations with lower energy in a vortex-free BEC are thoroughly
studied; the breathing or monopole mode with the azimuthal angular momentum $q_{\theta}=0$,
the dipole Kohn mode $q_{\theta}=1$, and the quadrupole mode $q_{\theta}=2$
for an axis-symmetric system \cite{dalfovoRMP}.

Recently, Bretin {\it et al}. \cite{bretin} have performed an experiment to examine the 
quadrupole modes with $q_{\theta}=\pm 2$ for a long-cigar shaped BEC with a vortex line,
observing the decay process.
Their results are summarized as follows:
(1) The one of the splitted quadrupole mode $q_{\theta}=-2$,
which rotates opposite to the vortex winding,
decays two-time faster than the other co-rotating quadrupole mode with $q_{\theta}=+2$.
(2) After the decay of $q_{\theta}=-2$ mode, 
there remains a density oscillation pattern along the long axis
($z$-axis) whose nodes are 7 or 8 within the length of the condensate.
The oscillation pattern is localized near the vortex core, seen in the radial direction profile.

Here we investigate the physical implication of these interesting observations,
by calculating the density-density response function based on the wave functions
and eigenvalues of the Bogoliubov-de Gennes equation for describing the collective
modes of Bosonic excitations.

%===================== Fig.1 ======================
\begin{figure}[b]
		(a) \includegraphics[width=2.9cm]{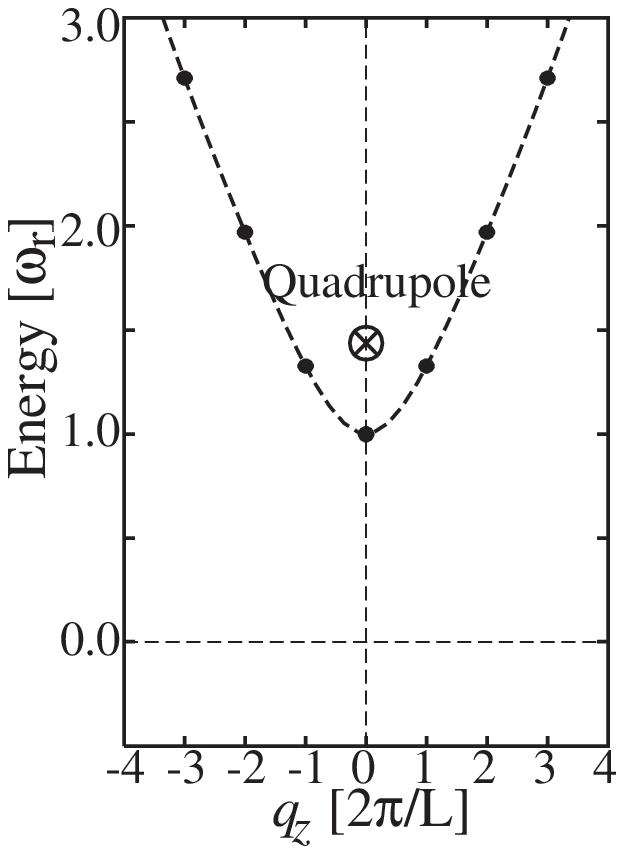}  
		(b) \includegraphics[width=2.9cm]{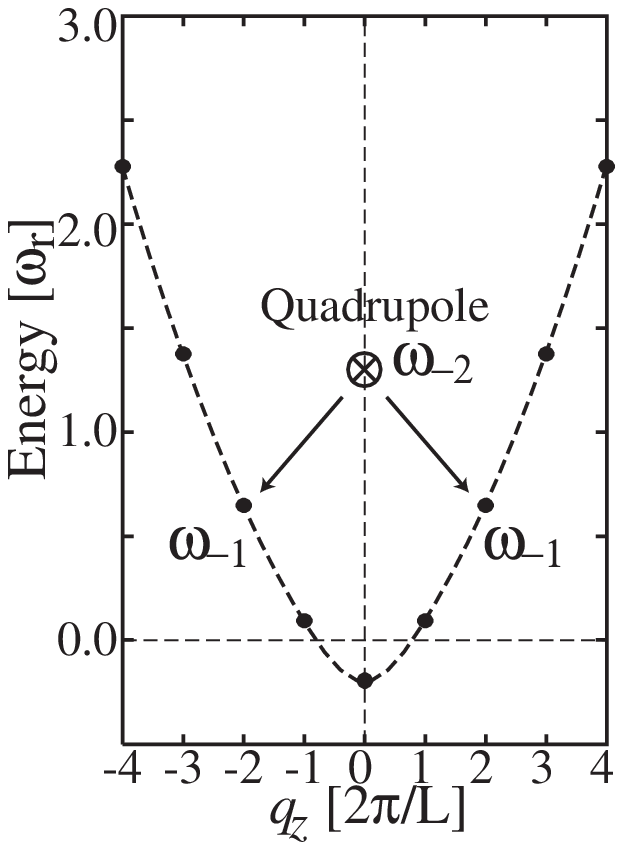}
\caption{Solid circles show the dispersion relations $\omega _{-1}(q_z)$ of
$(q_{\theta}, q_{r}) = (-1, 0)$ along $q_z$ in the case of 
the vortex-free state (a) and a single-vortex state (b).
$\otimes$ denotes $\omega _{-2}$.
}
\end{figure}
%===================== Fig.1 =====================/

Before going into detailed computation of a cylindrical system with a vortex,
we first give a clear physical picture for these phenomena which is fully justified microscopically 
later: 
At low temperatures, among the two possible decay channels,
the Beliaev process dominates over the remaining Landau process \cite{beliaev}. 
The counter-rotating quadrupole mode with the energy $\omega _{-2}$ 
and the angular momentum $q_{\theta}=-2$ 
(in units of the radial harmonic frequency $\omega _{r}$ and $\hbar = 1$ respectively) 
can decay into a pair of the dipole mode with $\omega _{-1}$,
conserving the energy $\omega _{-2}\!\rightarrow\!2\omega _{-1}$ and the angular momentum 
$(q_{\theta}=-2)\!\rightarrow\!(q_{\theta}=-1)+(q_{\theta}=-1)$.
However, as seen from Fig.~1 where a Beliaev process is depicted as a function
of the wave number $q_z$ along the $z$-axis,
the pair created dipole modes (see Fig.~1(b)) should have a finite $\tilde{q}_{z}$,
namely, 
$\omega _{-2} (q_{z}=0)
	\!\rightarrow\! \omega _{-1} (\tilde{q}_z) + \omega _{-1} (-\tilde{q}_z)$.
Note that in the absence of a vortex shown in Fig.~1(a) there is no Beliaev process because 
$2\omega _{1}(q_z) > \omega _2$, 
which is well-known, leaving only Landau decay active\cite{fedichev98L}.
Likewise as for the other comoving quadrupole mode with $q_{\theta}=+2$
there is no Beliaev process because $2\omega _{+1}>\omega _{2}$.
Only the Landau process, in which the quadrupole mode annihilates with the thermally
already excited modes into other modes,
is responsible for the decay.

Here it is central that the so-called anomalous mode $\omega _{-1}$ at $q_z=0$
has a negative values \cite{isoshima97,dodd,negns} relative to the condensate at the zero energy 
and that their wave function is radially localized at the core region whose wave length 
is an order of the coherence length.
Along the $z$-direction it oscillates sinusoidally with $\tilde{q}_{z}$.
These explain the above experimental facts (1) and (2) simultaneously.
In addition the decay time of $\omega _{+2}$ should be the same as in the vortex-free case,
which is indeed the case \cite{bretin}.

Let us now consider this novel phenomenon of the single-vortex condensate 
from the microscopic viewpoint.
The Hamiltonian in a rotating frame with the angular frequency ${\bf \Omega}_{{\rm rot}}$
is given by
%\begin{eqnarray}
$
 \hat{{\mathcal H}}
         = \int d {\bf r} \hat{\Psi}^{\dagger} ( {\bf r} )
             \left\{ h ({\bf r}) + \frac{g}{2} \hat{\Psi}^{\dag} ({\bf r})\hat{\Psi}({\bf r})\right\}
             \hat{\Psi} ( {\bf r} )  .
$
%\end{eqnarray} 
The creation and annihilation operators of the Bose particle are 
$\hat{\Psi}^{\dag}$ and $\hat{\Psi}$, which are decomposed into 
$\hat{\Psi} (\bf r) = \phi ({\bf r}) + \hat{\psi} ({\bf r})$.
The single-particle Hamiltonian is given as
$h({\bf r}) = - \frac{\hbar^2 \nabla^2}{2m} -\mu + V({\bf r}) 
                      - {\bf \Omega}_{{\rm rot}} \cdot ({\bf r} \times {\bf p})$
with the confining potential $V({\bf r})$ and 
the chemical potential $\mu$.
The last term in $\hat{{\mathcal H}}$ describes the interaction between the particles, 
which is classified into the resulting eight terms 
according to the noncondensate part $\hat{\psi}$,
by using the decomposition of the field operator obtained above \cite{FW}.

The quadratic Hamiltonian may be diagonalized by the usual Bogoliubov transformation,
\begin{eqnarray}
	\left(
	\begin{array}{c}
		\hat{\psi}  \\ \hat{\psi}^{\dag} 
	\end{array}
	\right)
	= \sum _{\bf q}
			\left(
			\begin{array}{cc}
				u_{\bf q}  & - v_{\bf q}^{\ast} \\
				-v_{\bf q} &   u_{\bf q}^{\ast}
			\end{array}
			\right) \left(
			\begin{array}{c}
				\eta _{\bf q} \\ \eta _{\bf q}^{\dag}
			\end{array} \right) 
	\equiv \sum _{\bf q}
			\hat{U}^{\dag} _{\bf q} 
			\left(
			\begin{array}{c}
				\eta _{\bf q} \\ \eta _{\bf q}^{\dag}
			\end{array} \right) .
\label{bogotrans}
\end{eqnarray}
This diagonalization leads to the following conditions.
First we impose the condition on the condensate wave function $\phi ({\bf r})$ as
%\begin{eqnarray}
$
	\left[ h({\bf r}) + g | \phi ({\bf r}) | ^2 \right] \phi ({\bf r}) = 0 ,
$
%\end{eqnarray}
which is the so-called Gross-Pitaevskii (GP) equation.
The Bogoliubov-de Gennes (BdG) equation for the quasiparticle is given 
in terms of the eigenfunctions $u_{\bf q}$ and $v_{\bf q}$ 
\begin{eqnarray}
	\begin{array}{ll}
		&\{h({\bf r}) + 2 g | \phi ({\bf r}) | ^2 \} u_{\bf q} ({\bf r}) 
			- g \phi ^2 ({\bf r}) v_{\bf q} ({\bf r}) 
			=   \varepsilon _{\bf q} u_{\bf q} ({\bf r}),  \\
		&\{h({\bf r}) + 2 g | \phi ({\bf r}) | ^2 \}v_{\bf q} ({\bf r}) 
			- g \phi ^{\ast 2} ({\bf r}) u_{\bf q} ({\bf r}) 
			= - \varepsilon _{\bf q} v_{\bf q} ({\bf r}),
	\end{array}
\label{eq:BdG}
\end{eqnarray}
\noindent
where ${\bf q}$ is the quantum number for the eigenstate.
It is well known that the eigenvalues $\varepsilon _{\bf q}$ correspond to 
the normal modes of the condensate and the values are good agreement with experiment
for systems with negligible thermal cloud \cite{dalfovoRMP}.

We consider the collective mode by the linear response theory \cite{FW}.
In the presence of an external field coupled to the density 
$\langle \hat{\Psi}^{\dag} \hat{\Psi} \rangle$,
the linear response is characterized by a retarded correlation function
$D^R({\bf r}{\bf r}',t-t')
	=-i\langle[\hat{\Psi}^{\dag}({\bf r}t)\hat{\Psi}({\bf r}t) , 
						\hat{\Psi}^{\dag}({\bf r}'t')\hat{\Psi}({\bf r}'t')]\rangle \theta (t-t')$,
where $\langle \cdots \rangle$ denotes the thermal average.
In order to calculate $D^R$ directly, 
it is convenient to introduce a Matsubara correlation function 
${\mathcal D}({\bf r}{\bf r}',\tau-\tau')
	=-\langle [\hat{\Psi}^{\dag}({\bf r}\tau)\hat{\Psi}({\bf r}\tau)  
						\hat{\Psi}^{\dag}({\bf r}'\tau')\hat{\Psi}({\bf r}'\tau')]\rangle$.
The Fourier coefficient ${\mathcal D}({\bf r}{\bf r}',i\Omega _n)$ can be related to 
the retarded correlation function, $D^R({\bf r}{\bf r}',\omega)$,
by using the analytic continuation $i\Omega _n \rightarrow \omega + i\eta$.
Here $\eta$ is a positive infinitesimal constant
and we use $\eta =0.005\omega _r$ in our calculation.
In the dilute Bose system, the Matsubara correlation function is characterized 
by a matrix 
\begin{eqnarray}
	{\mathcal D}({\bf r}{\bf r}',i\Omega _n)
	\simeq  - \left(
				\begin{array}{cc}
					\phi ^{\ast} ({\bf r}), & \phi ({\bf r})
				\end{array}
			  \right)
			\hat{{\mathcal G}} ({\bf r}{\bf r}', i\Omega _n)
			\left(
				\begin{array}{c}
					\phi ({\bf r}') \\ \phi ^{\ast} ({\bf r}')
				\end{array}
			  \right) ,
\end{eqnarray}
where $\hat{\mathcal G}$ is a $2\times 2$ matrix renormalized Green's function defined as 
$\hat{\mathcal G}({\bf r}\tau, {\bf r}'\tau')=-\langle T_{\tau}[ \hat{{\mathcal A}}({\bf r}\tau) 
																\hat{{\mathcal A}}^{\dag}({\bf r}'\tau')] \rangle$
with the matrix operator $\hat{{\mathcal A}}^{\dag}=(\hat{\psi}^{\dag}, \hat{\psi})$.
Using the Beliaev-Dyson equation, we have 
$\hat{\mathcal G} = (\hat{\mathcal G}_{0}^{-1} - \hat{\Sigma})^{-1}$ \cite{beliaev,FW}
with the bare Green's function $\hat{\mathcal G}_{0}$
which is constructed from quasiparticles 
and the canonical transformation Eqs.~(\ref{bogotrans}) and (\ref{eq:BdG}).

The self energy is expressed, by the second order perturbation theory on $g$,
as $\hat{\Sigma}=\hat{\Sigma}^{(1)}+\hat{\Sigma}^{(2)}$.
The first order term $\hat{\Sigma}^{(1)}$, which contributes to the energy shift, is
\begin{eqnarray}
	\hat{\Sigma}^{(1)} ({\bf q}_1 {\bf q}'_1, i\Omega _n)
	= g \int d{\bf r} \hat{U}^{\dag} _{{\bf q}_1} ({\bf r})
	\left(
		\begin{array}{cc}
			2\rho ({\bf r}) & \kappa ({\bf r}) \\
			\kappa ^{\ast} ({\bf r}) & 2\rho ({\bf r})
		\end{array}
	\right) \hat{U}_{{\bf q}_1} ({\bf r}),
\end{eqnarray} 
where the noncondensate density 
$\rho ({\bf r}) = \sum _{\bf q} [|u_{\bf q}({\bf r})|^2 f(\varepsilon _{\bf q})
														+|v_{\bf q}({\bf r})|^2\{f(\varepsilon _{\bf q})+1\} ]
$, the anomalous average 			
$\kappa ({\bf r}) = - \sum _{\bf q} u_{\bf q} ({\bf r})v_{\bf q}^{\ast}({\bf r})
										\{ 2f(\varepsilon _{\bf q}) + 1 \}$,
and the Bose function $f(\varepsilon _{{\bf q}})=[\exp(\beta \varepsilon _{\bf q})-1]^{-1}$
with $\beta = 1/k_{B}T$.
The second order term $\hat{\Sigma}^{(2)}$ is given by
\begin{widetext}
\vspace{-0.6cm}
\begin{eqnarray}
	\hat{\Sigma}^{(2)} ({\bf q}_1 {\bf q}'_1, i\Omega _n)
	&=& g^2 \sum _{{\bf q}, {\bf q}'}
			\left(
				\begin{array}{c}
					A_{1, {\bf q}{\bf q}'}({\bf q}_1) \\ A_{2, {\bf q}{\bf q}'}({\bf q}_1)
				\end{array}
			\right)
			\left(
				\begin{array}{cc}
					A_{1, {\bf q}{\bf q}'}^{\ast}({\bf q}'_1) & A_{2, {\bf q}{\bf q}'}^{\ast}({\bf q}'_1)
				\end{array}
			\right) 
			\frac{f(\varepsilon _{{\bf q}'}) - f(\varepsilon _{{\bf q}})}
					{i\Omega _n - (\varepsilon _{\bf q} - \varepsilon _{{\bf q}'})} \nn \\
	& & + \frac{g^2}{2} \sum _{{\bf q}, {\bf q}'}
			\left(
				\begin{array}{c}
					B_{1, {\bf q}{\bf q}'}^{a}({\bf q}_1) \\ B_{2, {\bf q}{\bf q}'}^{a}({\bf q}_1)
				\end{array}
			\right)
			\left(
				\begin{array}{cc}
					B_{1, {\bf q}{\bf q}'}^{a\ast}({\bf q}'_1) & B_{2, {\bf q}{\bf q}'}^{a\ast}({\bf q}'_1)
				\end{array}
			\right) 
			\frac{1 + f(\varepsilon _{{\bf q}}) + f(\varepsilon _{{\bf q}'})}
					{- i\Omega _n - (\varepsilon _{\bf q} + \varepsilon _{{\bf q}'})} \nn \\
	& & + \frac{g^2}{2} \sum _{{\bf q}, {\bf q}'}
			\left(
				\begin{array}{c}
					B_{1, {\bf q}{\bf q}'}^{b}({\bf q}_1) \\ B_{2, {\bf q}{\bf q}'}^{b}({\bf q}_1)
				\end{array}
			\right)
			\left(
				\begin{array}{cc}
					B_{1, {\bf q}{\bf q}'}^{b\ast}({\bf q}'_1) & B_{2, {\bf q}{\bf q}'}^{b\ast}({\bf q}'_1)
				\end{array}
			\right) 
			\frac{1 + f(\varepsilon _{{\bf q}}) + f(\varepsilon _{{\bf q}'})}
					{i\Omega _n - (\varepsilon _{\bf q} + \varepsilon _{{\bf q}'})} ,
\label{eq:self}
\end{eqnarray}
\end{widetext}
which determines the excitation and its damping.
In Eq.~(\ref{eq:self}), the first term describes Landau processes:
${\bf q}'_1 + {\bf q}' \rightarrow {\bf q} \rightarrow {\bf q}_1 + {\bf q}'$,
while the second and third terms correspond to Beliaev processes: 
${\bf q}'_1 \rightarrow {\bf q}+{\bf q}' \rightarrow {\bf q}_1$.
The matrix elements $A$ and $B$ are the overlap integrals of the condensate 
and the excitation amplitudes.
The frequency spectrum of the collective modes is given as 
\begin{eqnarray}								
S (\omega) = - \int d {\bf r} \int d {\bf r}^{'} 
                               F_{Q_{\theta}} ^{\ast} ({\bf r}) {\rm Im} 
							   D^R({\bf r}{\bf r}',\omega) F_{Q_{\theta}} ({\bf r}^{'}) .
\end{eqnarray}
In the case of the surface mode ($Q_{r}=0$), 
the general excitation operator is given as 
$F_{Q_{\theta}}=r^{|Q_{\theta}|} e^{iQ_{\theta}\theta}$ \cite{stringari96}.

We take up the quadrupole excitation experiment on $^{87}$Rb atoms
by Bretin {\it et al}.~\cite{bretin}.
Assuming their long-cigar system as a cylinder, 
we introduce a cylindrical coordinate: ${\bf r}=(r, \theta, z)$.
The quantum number ${\bf q}=(q_{\theta}, q_{z}, q_{r})$  
may take the following values:
$q_{\theta}=0, \pm 1, \pm 2, ...$, $q_z=0, \pm 2\pi /L, \pm 4\pi /L,...$, and $q_r=0,1,2,...$, 
where $L$ is the period of the length along the $z$-axis \cite{isoshima97} and we take $L=15\mu$m.
The linear density along the $z$-axis can be estimated as $n_z=6 \times 10^{9}/{\rm m}$
with the radial trap frequency $\omega _r /2\pi = 97.3$Hz ($T_c\sim 200$nK).
At finite temperatures, following the procedure by Rusch {\it et al}. \cite{rusch}, 
we use the chemical potential $\mu$ to fix the total number.

In Fig.~2(a) we show $S(\omega)$ in the frequency region
near the quadrupole excitations $\omega _{\pm 2}$,
where we used the parameters appropriate for the experiment by Bretin {\it et al}. \cite{bretin}.
The two resonance peaks are seen from it.
The peak around $\omega = 1.55 \omega _r$ corresponds to the 
main resonance of the quadrupole mode $q_{\theta}=+2$,
while the other mode $q_{\theta}=-2$ has a peak around $\omega = 1.3\omega _r$.
These resonance frequencies correspond to the observations 
($\omega _{+2}/2\pi=159.5\pm 1.0{\rm Hz}\simeq1.6 \omega _r /2\pi$,
$\omega _{-2}/2\pi=116.8{\rm Hz}\simeq 1.2\omega _r/2\pi$).
It is evident from Fig.~2(a) that the resonance width is wider in $q_{\theta}=-2$
than that in $q_{\theta}=+2$ by a factor two.

%===================== Fig.2 ======================
\begin{figure}[t]
	(a) \includegraphics[width=7.5cm]{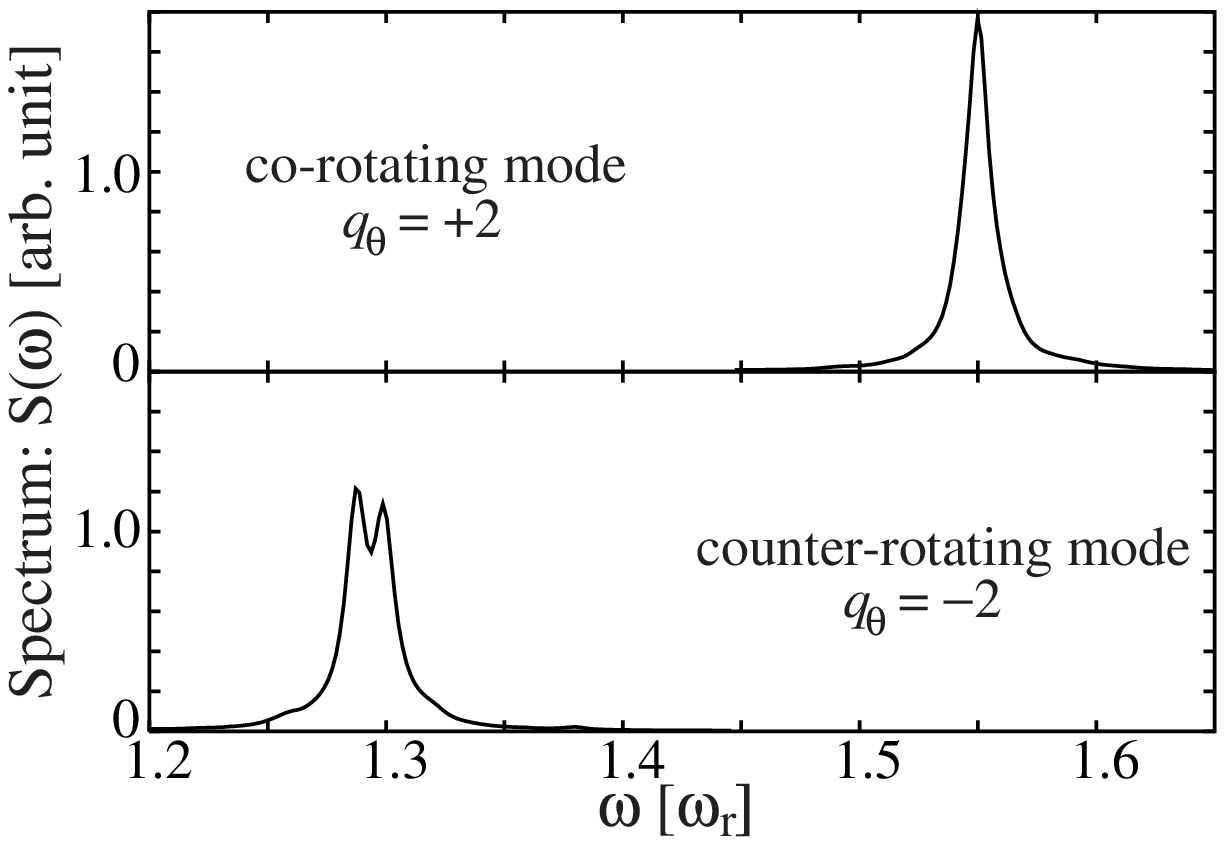} \\
	(b) \includegraphics[width=3.3cm]{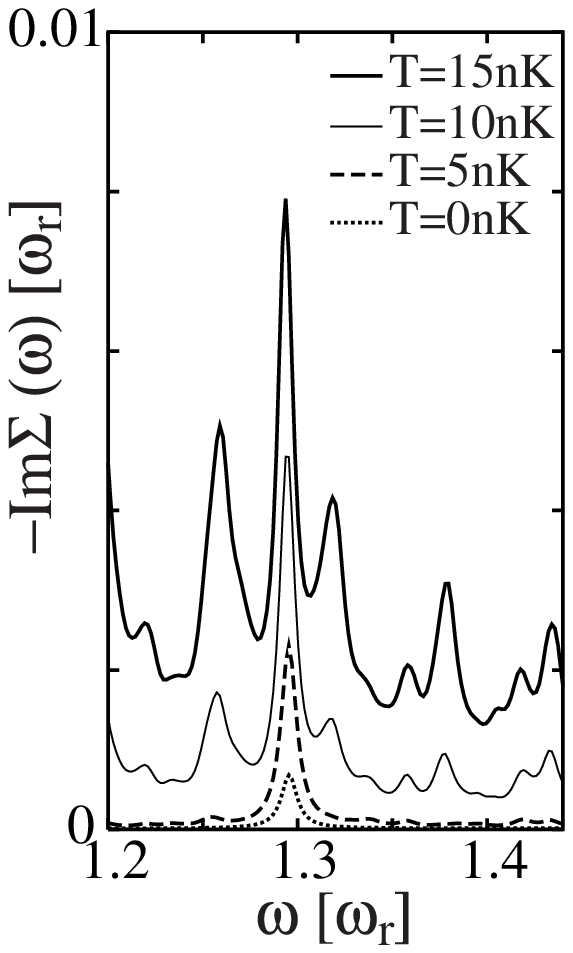} 
	(c) \includegraphics[width=3.3cm]{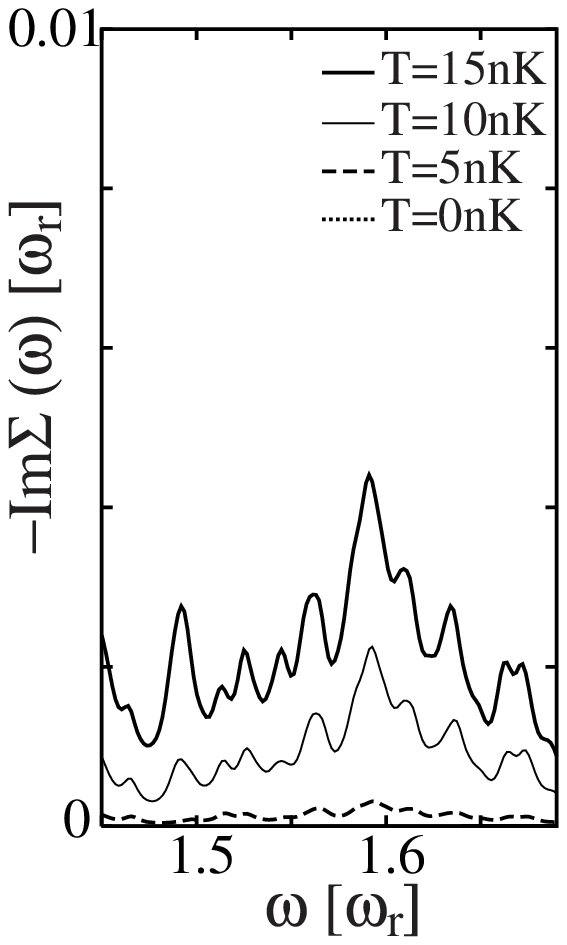}
\caption{
Spectrum of two quadrupole modes $q_{\theta}=\pm 2$ 
at $T=15$nK for $\eta = 0.005\omega _r$ (a).
The damping constant of the quadrupole mode $q_{\theta}=-2$ (b)
and $q_{\theta}=+2$ (c) at $T=0$, 5, 10, and 15nK.
The peak around $\omega = 1.3\omega _r$ corresponds to 
the Beliaev process $(q_{\theta} =-2,q_z =0) \rightarrow (-1, +2) + (-1, -2)$.
}
\end{figure}   
%===================== Fig.2 =====================/

Since the damping of the collective mode comes from  the imaginary part of the self energy
$\hat{\Sigma}^{(2)}$ in Eq.~(\ref{eq:self}) and the dominant matrix element is $\Sigma ^{(2)}_{11}$,
we discuss the damping constant defined as 
$\Gamma _{q_{\theta}}(\omega)
	=- {\rm Im} \Sigma ^{(2)}_{11}({\bf q},{\bf q}, \omega) | _{q_r = q_z =0}$.
The damping constant for $q_{\theta}=-2$ and $q_{\theta}=+2$ are,
respectively, shown in Fig.~2(b) and (c).
For the quadrupole mode $q_{\theta}=+2$, only the Landau process
$(+2, 0,0)+(q_{\theta}, q_z,q_r) \rightarrow (q_{\theta}+2, q_z,q'_r)$ is active 
as in the vortex-free case. Thus the damping constant for $q_{\theta}=+2$ is 
nearly same for both single-vortex and vortex-free cases.
This is exactly seen by Bretin {\it et al}. \cite{bretin}.
On the other hand, the Beliaev process exclusively dominates the damping of the $q_{\theta}=-2$
mode because, as mentioned before, ($-$2, 0, 0) can decay into the two dipole modes 
$\omega _{-1}$ by conserving the angular momentum and the total energy.
As is seen from Fig.~2(b), 
$\Gamma _{-2} (\omega\sim 1.3\omega _{r})$ for $q_{\theta}=-2$ 
at $T=0$ shows a single peak due to this particular Beliaev process
while $\Gamma _{+2} (\omega \sim 1.55 \omega _{r})$ 
for $q_{\theta}=+2$ shows no prominent peak (Fig.~2(c)).
As $T$ increases, the Landau processes becomes effective, giving rise to the satellite peaks
in addition to  the main peak as shown in Fig.~2(b).
We can see that at $T\sim 15$nK 
the ratio of two damping constant $\Gamma _{\pm 2}$ at each resonance
is $\Gamma _{-2}/\Gamma _{+2}\simeq 2$,
which agrees with the observation \cite{bretin}.
However, this ratio depends sensitively on the temperature;
As $T$ decreases, the Landau process quickly vanish,
leaving only the particular Beliaev process active.
This should be checked experimentally.

The created modes via the Beliaev process of the counter rotating quadrupole mode $q_{\theta}=-2$
can be identified to the so-called anomalous mode \cite{isoshima97,dodd}, 
or Kelvin mode characterized by the quantum
number $q_{\theta}=-1$, $q_r=0$, and $\pm \tilde{q}_z=4\pi/L$ as shown in Fig. 1(b).
This dipole mode is counter-rotating to the vortex flow direction.
Since this mode has the zero-relative angular momentum to the condensate,
the wave function $u(r,z)=u_{q_{\theta}=-1}(r)e^{iq_zz}$ does not vanish 
at the center $r=0$ and localized at the core while all other wave functions with $q_{\theta}\neq -1$
vanish at $r=0$.
In Fig.~3 we display the condensate $|\phi(r,z)|$ and the wave function of the Kelvin mode with  
$\tilde{q}_z = \pm 4\pi/L$, which is created by decaying of the counter-rotating quadrupole mode.
One can see that this anomalous mode with a negative eigenvalue at $q_z=0$,
which is  first identified theoretically Isoshima and Machida \cite{isoshima97} 
and Dodd {\it et al}. \cite{dodd},
is localized within the core region along the 
radial direction whose scale is the coherence length. 
The characteristic wave number $\tilde{q}_z=4\pi/L$ approximately corresponds to 
the observation \cite{bretin}.
These particular features, the localization around the core in the radial direction
and propagation along the vortex line direction, 
are exactly what Bretin {\it et al}. \cite{bretin} have detected.
Since there is no core localized mode other than this $q_{\theta}=-1$ mode \cite{isoshima97},
we conclude that a pair of these anomalous modes with $\tilde{q}_z$ and $-\tilde{q}_z$
are created. Thus the Kelvin wave, or the wave motion of the vortex line 
propagating along $z$-axis, is now identified and imaged by their experiment 
(see Fig.~3 in Ref. \cite{bretin}).

%===================== Fig.3 ======================
\begin{figure}[t]
	\begin{tabular}{cc}
		\includegraphics[width=3.9cm]{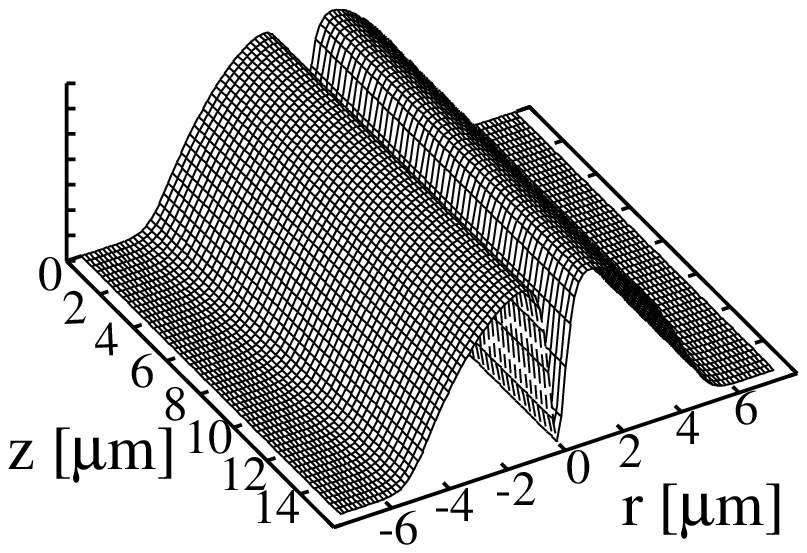}  &
		\includegraphics[width=3.9cm]{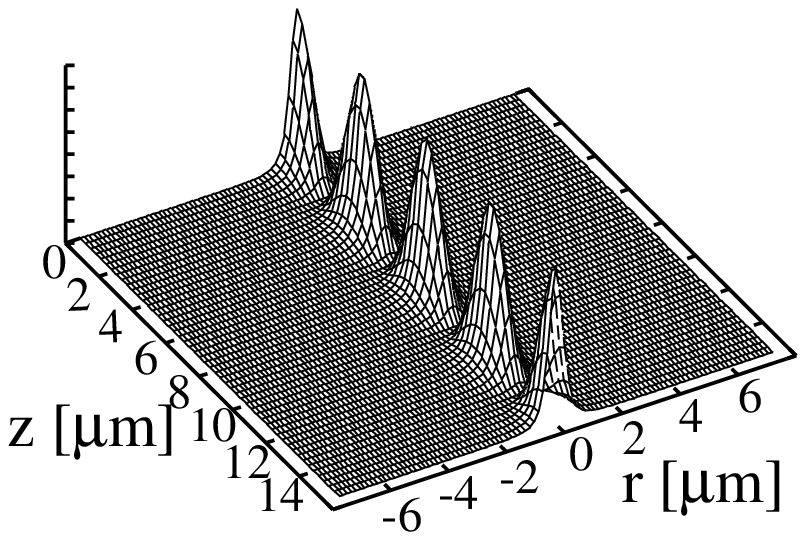} \\
		(a) & (b)
	\end{tabular}
\caption{
Spatial profiles of 
the condensate $|\phi (r, z)|$ (a)
and the Kelvin mode  $|u (r, z)| = | u_{q_{\theta} = -1} (r) \cos(\tilde{q}_z z) |$
with $\tilde{q}_{z} = \pm 4 \pi / L$ (b).
}
\end{figure}   
%===================== Fig.3 =====================/

In the following we consider several situations to help identifying 
the Kelvin mode: Let us first consider the external rotation effect on the decay process.
Under the external rotation frequency $\Omega _{{\rm rot}}$,
$\omega _{\pm 2}(\Omega _{{\rm rot}})=\omega _{\pm 2} \mp 2\Omega _{{\rm rot}}$
as is seen from Eq.~(\ref{eq:BdG}) \cite{fetterJP}.
Accordingly, the spectral response function $S(\omega)$ shown in Fig.~4
exhibits; (1) The two resonances switch their positions 
at around $\Omega _{{\rm rot}}\simeq 0.05\omega _r$, 
and (2) the resonance widths become comparable as $\Omega _{{\rm rot}}$ increases.
It is due to the suppression of the Beliaev decay,
because under the rotation the population of the possible modes decreases according to the rule 
$\omega _{q_{\theta}}(\Omega _{{\rm rot}})
	=\omega _{q_{\theta}} (0) - q_{\theta}\Omega _{{\rm rot}}$.

The finite temperature affects both Beliaev and Landau processes.
As increasing $T$, $\omega _{-1}(T)$ is known to be larger while $\omega _{-2}(T)$ is 
relatively independent of $T$ except the region near $T_c$ \cite{isoshimaT}.
Thus $\tilde{q}_z(T)$ becomes small with $T$ and above some critical temperature the 
Beliaev process $\omega _{-2}\rightarrow \omega _{-1}(\tilde{q}_z)+\omega _{-1}(-\tilde{q}_z)$
ceases to exist, unabling to create the Kelvin modes
as the decay channel of $\omega _{-2}$.
Generally the Landau process for both $\omega _{\pm 2}$ modes
becomes important as $T$ increases because thermally excited modes become more available.

In Bretin {\it et al}. experiment \cite{bretin} the counter-rotating $\omega _{-2}$
mode is used to excite the Kelvin mode via the Beliaev process.
It is also possible to use the monopole mode $\omega _0$ 
to create the Kelvin mode, namely 
$\omega _0\rightarrow \omega _{+1}(\tilde{q}_z)+\omega _{-1}(-\tilde{q}_z)$.
This provides yet another spectroscopic method to analyze the Kelvin mode.

In conclusion, we have demonstrated that the Kelvin mode as a propagation wave 
along a vortex line can be excited via Beliaev decay processes for 
the counter-rotating quadrupole mode.
It enables us to understand the experiment by Bretin {\it et al}. \cite{bretin} successfully,
namely, their resonance position and width,
and to predict the external rotation and temperature effects.
We have also shown that utilizing these decay channels provides a novel spectroscopic 
method for low-lying Bosonic excitations in a vortex,
in particular, the unexplored Kelvin mode.

%===================== Fig.4 ======================
\begin{figure}[t]
	\includegraphics[width=6.7cm]{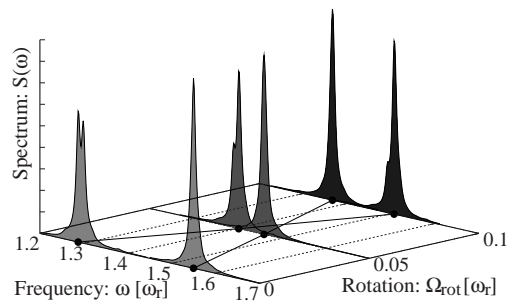} 
\caption{
The external rotation effect on spectrum of the co- and counter-rotating 
quadrupole modes at $15$nK.
}
\end{figure}   
%===================== Fig.4 =====================/

The authors thank P. Rosenbusch for communicating their results prior to publication.


\begin{thebibliography}{2}

\bibitem{dalfovoRMP}
F. Dalfovo {\it et al}., Rev. Mod. Phys. {\bf 71}, 463 (1997).

\bibitem{fetterJP}
A.L. Fetter and A.A. Svidzinsky, J. Phys.: Condens. Matter {\bf 13}, R135 (2001). 

\bibitem{matthews}
M.R. Matthews {\it et al}., Phys. Rev. Lett. {\bf 83}, 2498 (1999). 

\bibitem{madison}
K.W. Madison {\it et al}., Phys. Rev. Lett, {\bf 84}, 806 (2000).

\bibitem{leanhardt}
A.E. Leanhardt~{\it et~al}.,~Phys.~Rev.~Lett.~{\bf 89},~190403~(2002).


\bibitem{abo01}
J.R. Abo-Shaeer {\it et al}., Science {\bf 292}, 476 (2001).

\bibitem{sc}
In high $T_c$ cuprates the magnetic field induced vortex state 
is one of the focused themes.
See for example,
S. Sachdev and S.-C. Zhang, Science {\bf 295}, 452 (2002).

\bibitem{donnelly}
R.J. Donnelly, {\it Quantized Vortices in Helium II}
(Cambridge University Press, Cambridge, 1991).

\bibitem{bretin}
V. Bretin {\it et al}., cond-mat/0211101.

\bibitem{beliaev}
S.T. Beliaev, Sov. Phys. JETP {\bf 7}, 289 (1958).

\bibitem{fedichev98L}
P.O. Fedichev {\it et al}., Phys. Rev. Lett. {\bf 80}, 2269 (1998).

\bibitem{isoshima97}
T. Isoshima and K. Machida, J. Phys. Soc. Jpn. {\bf 66}, 3502 (1997).

\bibitem{dodd}
R.J. Dodd {\it et al}., Phys. Rev. A {\bf 56}, 587 (1997). 

\bibitem{negns}
It is generally accepted that it is negative with the positive norm 
conserved wave function.
Virtanen {\it et al}. 
[S.M.M. Virtanen {\it et al}., Phys. Rev. Lett. {\bf 86}, 2704 (2001)] 
claim that it could become almost zero even at zero-temperature. 
This point is still under discussion. 
See for review, Ref.~\cite{fetterJP}.

\bibitem{rusch}
M. Rusch {\it et al}.,  Phys. Rev. Lett. {\bf 85}, 4844 (2000);
S.A. Morgan, J. Phys. B {\bf 33}, 3825 (2000).

\bibitem{FW}
A.L. Fetter and J.D. Walecka, 
{\it Quantum Theory of Many-Particle Systems}
(McGraw-Hill, New York, 1971).

\bibitem{stringari96}
S. Stringari, Phys. Rev. Lett. {\bf 77}, 2360 (1996).

\bibitem{isoshimaT}
T. Isoshima~and~K. Machida,~Phys.~Rev.~A~{\bf 59},~2203~(1999).

\end{thebibliography}
\end{document}